# RAPID DEPRESSURIZATIONS: CAN THEY LEAD TO IRREVERSIBLE DAMAGE?

Pierre Berest and Hippolyte Djizanne
LMS, Ecole Polytechnique, Palaiseau, France

Benoît Brouard, Brouard Consulting, Paris, France

Grégoire Hévin, Storengy, Bois Colombes, France

**Abstract**

Rapid gas depressurization leads to gas cooling that is followed by slow gas warming when the cavern is kept idle. The decrease in the temperature of gas depends upon the relative withdrawal rate (in %/day), and cavern size and shape. Gas cooling may result in the onset of tensile stresses at cavern walls and roofs that may generate fractures or cracks. However, in most cases, the depth of penetration of these fractures is small, and they are perpendicular to the cavern wall. The distance between two parallel fractures becomes larger when fractures penetrate deeper into the rock mass, as some fractures stop growing. Fractures form a polygonal pattern. Salt slabs are created, with boundaries formed by the opened fractures. As long as the depth of penetration of the fracture remains small, these slabs remain strongly bonded to the rock mass, and it is believed that, in many cases, their weights are not large enough to allow them to break off the cavern wall.

## 1. INTRODUCTION

Gas storage caverns were developed mainly for seasonal storage, with one or a few cycles per year and a moderate gas-production rate between the maximum and minimum operation pressure. However, the needs of energy traders are prompting change toward more aggressive operating modes. Typically, high-deliverability caverns can be emptied in 10 days and refilled in 30 days or less (Spreckels and Crotogino, 2002). Maximum pressure-drop rates, which used to be in the range 1-2 MPa/day (150-200 psi/day) range, are expected to become faster (Gilhaus, 2007).

At the same time, Compressed Air Energy Storage (CAES) is experiencing a rise in interest, as it can be used as buffer energy storage in support of intermittent sources of renewable energy, such as wind mills. CAES facilities are designed to deliver full-power capacity in a very short time period.

Both types of facilities (CAES and High-Frequency Cycled Gas-Storage Cavern, HFCGSC) imply high gas-production rates and multiple yearly pressure cycles. However, two differences are worth noting: (1) cycles in a CAES are more frequent than in an HFCGSC, up to one cycle per day; and (2) the gap between maximum and minimum pressure in a CAES is smaller. [The Huntorf and McIntosh CAES facilities are operated between 7 and 4.3 MPa, or 1000 and 600 psi (Crotogino et al., 2001)].

This cycled mode of operation of solution-mined caverns raises questions regarding frequently repeated, extreme, mechanical and thermal loading. The standard approach used for designing classical gas storage (including salt constitutive-creep laws, laboratory testing procedures, numerical computations and stability criteria) must be revised in this new context. In this note, we discuss the thermal and mechanical consequences of a fast pressure drop (during gas withdrawal), followed by a period of time during which the cavern is kept at a standstill.

Fast pressure drop can generate drastic gas cooling (by several dozens of °C or °F). At the cavern wall, salt is much colder than it is in the rock mass, and large tensile stresses are generated. They are as large as 1 MPa per °C of temperature change (or 80 psi par °F of temperature change). Depending on the initial (before depressurization) stress distribution, fractures may open. It is feared that fractures lead to dramatic spalling and loss of cavern tightness. This paper provides proof that such risks exist, but must not be overestimated.

## 2. TWO EXTREME EXAMPLES

### 2.1. Warming

A case of intense rock warming was described by Lee et al. (1982). Hot (315 °C) diesel-exhaust gases from an underground power plant were circulated in a horizontal drift in a granitic rock formation at a 150-m (450 ft) depth at North Bay, Ontario, Canada. After several months, the cross-sectional area of the drift had increased by 100% due to intense spalling from the roof and walls.

> "*The fracture surfaces appeared to ignore the structure in the rock and to be determined mainly by the pattern of the thermal stress.*" (Lee et al., 1982, p. 964) (see Figure 1).
>
> *"A similar mode of surficial spalling was observed in a smaller test passage (0.76 m by 0.76 m in cross-section and 3 m long) [...] excavated for the purpose of experimental study [...] Test runs showed that spalling occurred at a surface temperature rise as low as 61 °C. One run was continued for 8-1/2 hours and a mass of spall was produced as shown in [the] figure [1]."* (Lee et al., 1982, p. 964.)

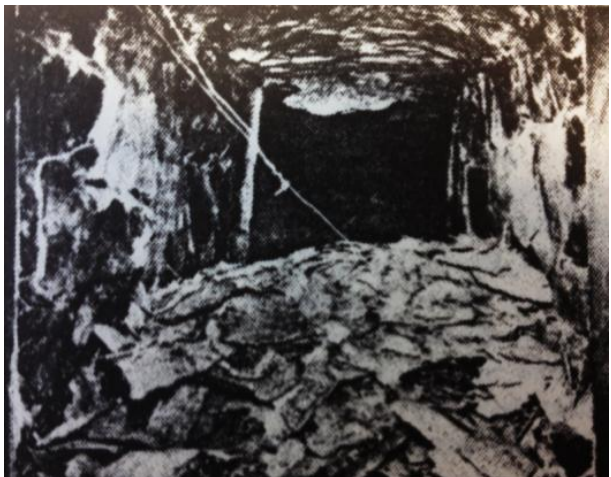

**Figure 1.** Test passage at North Bay, Ontario, Canada (Lee et al., 1982).

## 2.2. Cooling

Dreyer (1977) leached out a small cavern below a mine drift at a 600-m (1800-ft) depth. The cavern height and diameter were 2 m (6 ft) and 1 m (3 ft), respectively. The cavern was filled with liquid nitrogen whose temperature was 77 K (-196 °C or -320°F). At the end of the test, four fractures were observed, as shown on Figure 2.

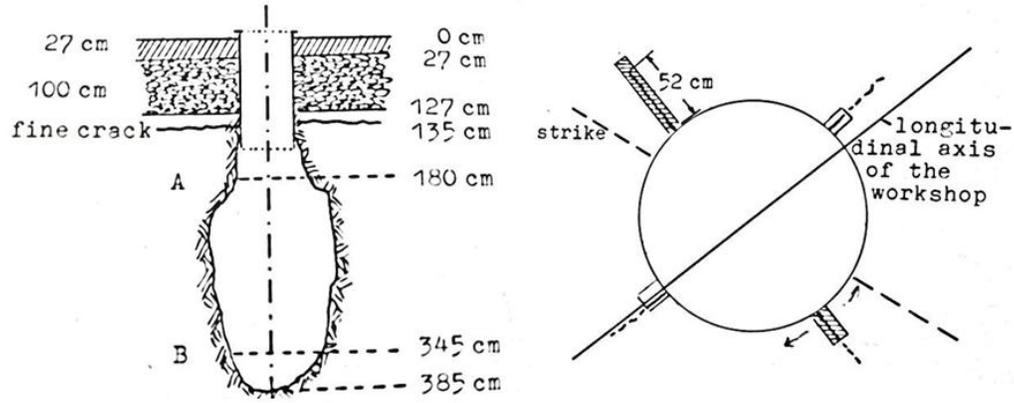

**Figure 2.** Fracture development in a small cavern leached out below a mine drift (Dreyer, 1977).

These two somewhat extreme cases (Fluid temperature at the cavern wall was increased by 300°C or decreased by -200°C, respectively.) prove that warming and cooling have no symmetrical effects. Warming leads to spalling; cooling leads to fracturing. This will be discussed in Section 5.

## 3. LESSONS DRAWN FROM ACTUAL CAVERNS

Several caverns experience spalling, sluffing, break-outs and fast creep closure. In most cases, these caverns were deep gas-storage caverns in which gas pressure was low when the gas inventory was small. Were these anomalous behaviors related to rapid depressurization and fluid cooling?

### 3.1. Kiel (Germany)

The Kiel 101 cavern (Figure 3) had been leached out at a depth between 1305 m and 1400 m (Röhr, 1974). The amount of insolubles in the salt formation was high, resulting in a large sump at the cavern bottom. The cavern was to be operated as a gas storage. After leaching was completed, a test was performed. A submersible pump was lowered to the bottom of the well. *"Starting about 1 November 1967, the pressure at the roof of the cavity was lowered from 15.6 MPa to practically zero by pumping the brine out of the access well"* (Baar, 1977, p. 147). In fact, when cavern pressure was lowered from 13.1 MPa to 6.5 MPa in 5 days, the cavern roof broke (Figure 3). A sonar survey was performed at the end of the testing period. The volume accessible to sonar had decreased from 36,600 $m^3$ to 32,100 $m^3$. An additional loss of 1900 $m^3$ was observed 5 months later. Many factors could have played a role in this evolution — e.g., a heterogeneous salt formation, a flat roof and a rapid depressurization (more than 1 MPa/day). However, fluid cooling cannot be a factor (during depressurization, the brine temperature drops by 0.03°C/MPa, a very small figure.)

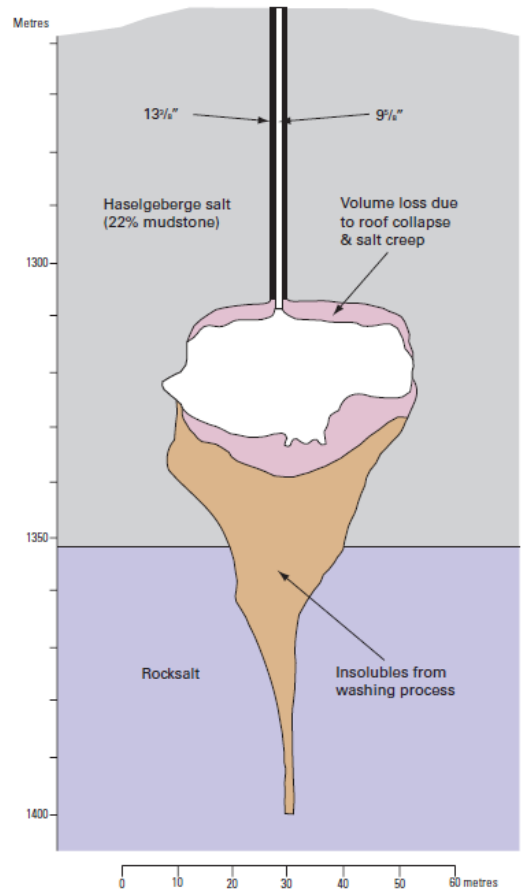

**Figure 3.** Kiel Cavern, Germany (British Geological Survey, 2008).

### 3.2 Eminence Salt Dome (Mississippi)

A somewhat similar example is provided by Cavern #1 of the Eminence Salt Dome, Mississippi (Baar, 1977; Serata and Cundey, 1979; Coates et al., 1983). Its depth was between 1725 m and 2000 m, and the geostatic pressure was $P_\infty = 40$ MPa (5600 psi) or so. The first gas injection was completed by mid-1970. Gas pressure was lowered to 6 MPa (840 psi) and after October 1970, gas pressure was kept constant for 2 months. After this period, it was increased to 28 MPa. A similar pressure cycle was performed in 1971, for which large volume losses were suspected. By April 1972, the cavern bottom had raised by 36 m. A sonar survey was performed and showed that the cavern volume loss was 40%. It was noticed (Figure 4), however, that the cavern roof seemed to have remained intact. It is difficult to assert whether, in addition to cavern loss, sluffing and spalling took place. It is obvious that a large part of the volume loss must be attributed to bottom heave. Berest et al., 1986, suggest that, in addition to possible block falls, this non-symmetrical behavior can be explained at least partly by larger overburden pressure and warmer temperature in the lower part of the rock mass. Note, also, that the gas depressurization was not overly fast in this cavern

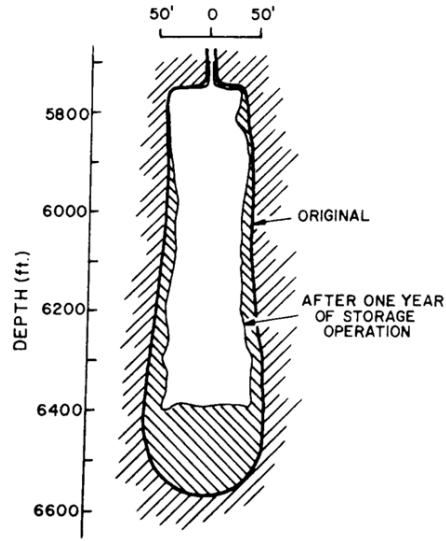

**Figure 4.** Cavern #1, Eminence Salt Dome, Mississippi (Serata and Cundey, 1979).

### 3.3. Tersanne, Drôme, France

TE02 is a cavern belonging to the Tersanne (France) gas-storage facility operated by Storengy (Hévin et al., 2007). Cavern depth is 1400 to 1550 m, a depth at which the overburden pressure is $P_\infty = 32$ MPa (4500 psi) or so. This cavern was leached out from November 1968 to February 1970. At this date, according to Boucly and Legreneur, 1980, the "sonar" volume was $V = 91600 \pm 4300$ m$^3$. The cavern was filled with gas and operated as a "seasonal" storage. It experienced 9 pressure cycles between 22 MPa and 8 MPa during the 1970-1979 period. The measured cavern gas temperature fluctuated between 32°C (90°F) and 69°C (155°F).

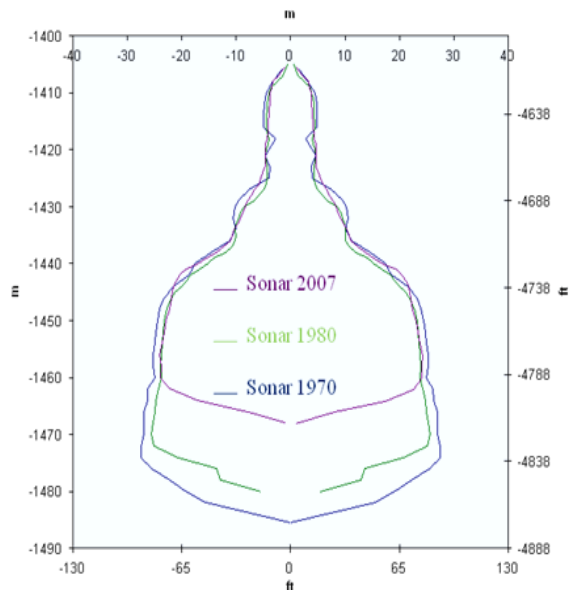

**Figure 5.** TE02 cavern shape evolution (Source : Grégoire Hévin, Storengy).

On July 1979, it became clear that cavern had lost significant volume; the cavern bottom had risen by 5.8 m. It was decided to fill the cavern with brine and perform a sonar survey. Cavern volume had decreased to $V =$ 58 200 ± 2300 m$^3$, a 1/3 loss reduction in a 9-year-long period.

> *In order to explain the gas-brine interface rising it is not possible to exclude the fall of the blocks following the failure of the overhanging banks visible on the February 1970 reading; the January 1980 reading has a much smoother appearance (however, the more rapid creep for salt than for anhydrite can also be brought into consideration, this would also give a smoother appearance on the reading.*), Boucly and Legreneur, 1980, p.256.

### 3.4. Huntorf (Germany)

Crotogino et al., 2001, described "*more than 20 years of successful operation*" of the CAES storage at Huntorf (Germany). Two caverns with depths between 650 m and 800 m were operated. Air pressure varied between 4.3 and 7 MPa (600 and 1000 psi). The fastest depressurization rate was 5 MPa/hr (700 psi/hr). Between 1978 and 1986, there were 200 starts/yr between 1978 and 1986, and less than100 starts/yr in the 1990s.

Quast, 1983, compared two vertical cross-sections of the NK1 cavern, which had been run in 1976 (filled with brine) and four years later in 1980 (filled with gas). His opinion was that the (laser) 1980 sonar survey correctly reflects the actual shape of the cavern. Salt "curtains", still present in 1976 when the cavern was filled with brine, broke and fell to cavern bottom during the first gas injection (during which the cavern bottom raised by 16 m). Later on, the depth of the brine-air interface remained almost constant, and Crotogino et al., 2001, state that "*evaluations over the whole operational period shows practically no changes that can be attributed to roof falls.*" (p. 356). Although laser sonar could not be performed routinely,

> "*when cavern NK1 was expanded to atmospheric pressure at the beginning of 2001, a survey was possible with a heated laser tool […]. The evaluation of this survey after over 20 years of operation showed practically no deviation compared to the original conditions*" (p.356).

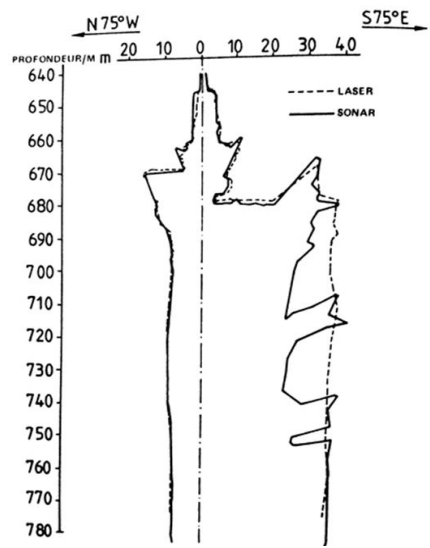

**Figure 6.** Huntorf Cavern: Comparison between a 1976 sonar survey (cavern filled with brine) and a 1980 sonar survey (cavern filled with gas) (Quast, 1983).

## 3.5. Lille-Torup, Danemark

Rokahr et al., 2007, described a remarkable rock mechanics test performed in a cavern of the Lille-Torup (Denmark) gas-storage facility. The average cavern depth was 1460 m. During the test, the cavern pressure dropped from 21 MPa (2940 psi) to 6 MPa (840 psi) in three months. Two high-resolution (laser) sonar surveys were performed on January 16, 2005 and March 31, 2005 (Figure 7), when the gas pressure was 14 MPa (1960 psi) and 6 MPa (840 psi), respectively. Cavern cross section increased as is clearly visible on Figure 7: *"The enlargement of the cavern in this sector is interpreted as a consequence of local failure of the rock salt at the cavern wall, that means that spalling had occured locally."* (Rokahr et al., p.137). The authors assess that the overall volume of rock that fell to cavern bottom during the test was 1000 to 2000 $m^3$.

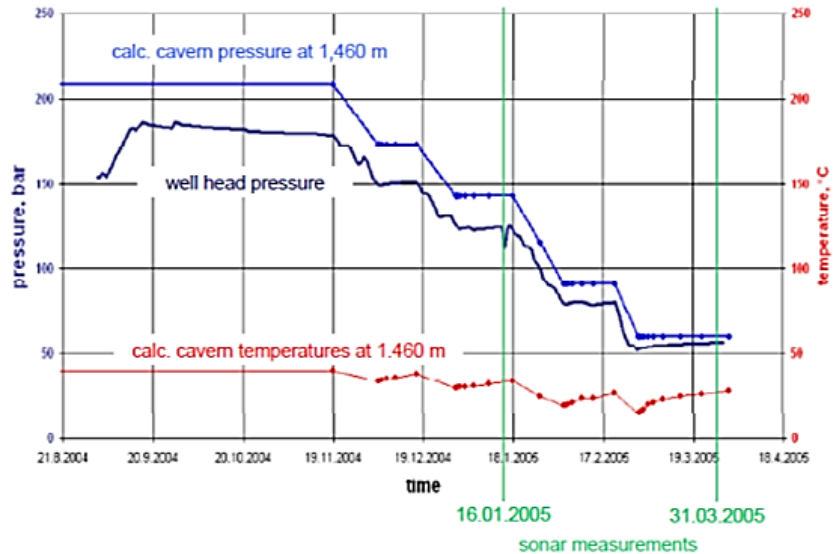

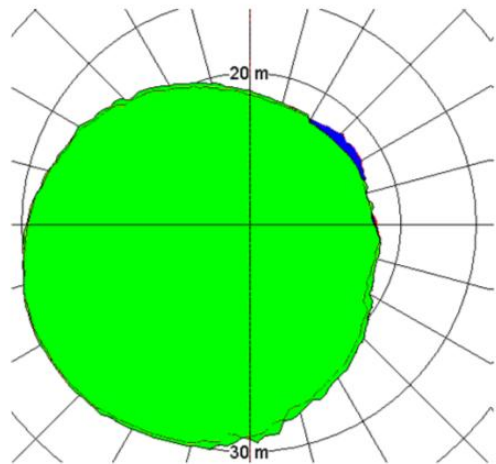

**Figure 7.** Horizontal cross-sections at a 1460-m depth. Two sonar surveys were performed on January 16, 2005 and March 31, 2005, over which time the cavern pressure dropped from 14 MPa to 6 MPa. The effects of spalling can be observed in the N-E part of the cross-section (after Rokahr et al., 2007).

### 3.6. Markham (Texas)

Cole, 2002, describes two gas caverns of the Markham storage field in Texas. These caverns were cycled 8 to 10 times per year. Withdrawal rates were 3-4 times faster than injection rates. Cavern tops are at 3450 ft (1050 m) and 3531 ft (1075 m), and the original cavern heights were 1739 ft (530 m) and 2284 ft (696 m), respectively. The first gas injections were in June 1992 (Cavern #2) and October 1995 (Cavern #5). Cavern operation resulted in cavern-volume loss, and salt sluffing from the walls and roof of the cavern. Brine-gas interface surveys were run every year. As of January 2002, the caverns had lost 292 ft (89 m) and 419 ft (128 m) of their depths. A "material balance" test also had been performed in 2001 to assess the cavern "free" volume, and natural gas sonars were run in January 2002. Comparison of these various methods suggests that the cavern creep closure was approximately 5% of the initial volume; however, a similar volume of salt had fallen from the walls to the bottom of the cavern. Comparison of sonar profiles (Figure 8) proved that "*no large sections of salt from one area of the caverns fell to the bottom creating the fill ... the salt was removed in thin layers over large areas of the caverns ... horizontal cavern closure occurred ... thus closing in part of the area created by the salt falling.*" (Cole, 2002, p. 82).

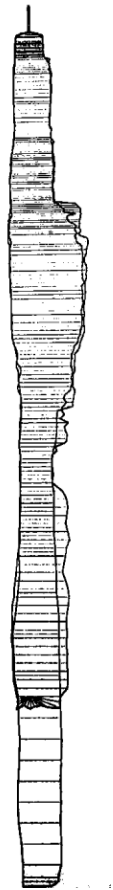

**Figure 8.** Markham Cavern #5. Vertical Profile Overlay of Sonars performed in May 1995 and January 2002. Note bottom heave and possible break-outs at cavern wall (Cole, 2002).

## 3.7. Conclusions

Data are not unequivocal. Large creep closures were observed at Tersanne and Eminence. Some spalling also took place, but it is suspected to have been small. Depressurization rates were relatively slow, and gas cooling does not seem to have been an important factor. Overhanging blocks fell down during the first gas injection at Huntorf; this may be attributed to the reduction in Archimedes thrust, when fluid density drastically drops after brine withdrawal. No, or small, further spalling was observed despite fast depressurization rates. Block falls certainly occurred at Lille-Torup, but the gas-temperature drop was small, as the pressure-drop rate was 5 MPa/month (700 psi/month) or so. Roof fall at Kiel certainly can be attributed to a large and fast pressure drop; however temperature changes in a *brine*-filled cavern cannot be a factor. Markham is a remarkable case, as many pressure cycles were performed (8-10/year), and both cavern creep closure and wall sluffing were observed.

## 4. TEMPERATURE CHANGES IN A GAS CAVERN

### 4.1. Convection in a gas-filled cavern

In a gas cavern, pressure is almost uniform, as gas density is small (50 to 200 kg/m$^3$). Temperature distribution is a slightly more complicated problem. In a salt formation, rock thermal conductivity is $K_R$ = 5 to 6 W/m/°C, and the geothermal gradient typically is $G$ = 1.5 to 1.8 °C/100 m. Temperature is warmer at the cavern bottom, gas is lighter, and it can be expected that gas effectively is stirred by thermal convection. In fact, natural convection only occurs when the geothermal gradient is larger than a certain threshold, the *adiabatic* gradient ($G_{ad}$). For a dry gas, $G_{ad} = g/C_P$ where $g$ = 10 m/s$^2$ is the gravity acceleration, and $C_P$ is the specific heat of gas (when pressure is kept constant). For natural gas, $C_P$ = 2345 J/kg/°C and for air $C_P$ = 1000 J/kg/°C. In other words $G > G_{ad}$ and, in principle, convection must appear in any cavern. However, $G$ is not much larger than $G_{ad}$, and convection can be impeded in some cases — for instance, when warm air is injected at the cavern top, when cold brine is left at the cavern bottom at the end of the leaching period, or when gas is cooled to a temperature colder than the temperature of the brine left at the cavern bottom. Examples can be found in Fosse and Røvang, 1998, Kneer et al, 2003, Klafki et al, 2003, Krieter et al. 1998, Quast, 1983 and Skaug et al., 2010. However, when thermal convection develops, it is extremely effective. Dimensional analysis shows that convection is governed by the Prandtl number ($Pr$ = $v/k$, $v$ = kinematic viscosity, $k$ = gas thermal diffusivity) and the Grashof number ($Gr = g\alpha Ga^4 / v^2$, where $a$ is the cavern characteristic length). The Grashof number is quite high in a large cavern, and turbulent convection develops. Note that in a *brine*-filled cavern $G_{ad} = \alpha Tg/C_P = 0.1$ °C/100m: onset of convection is certain.

### 4.2. Heat Balance Equation

When it is assumed that gas temperature, $T$, and pressure, $P$, are almost uniform throughout the entire cavern, the heat balance equation can be written (ATG, 1986; this equation is at the base of the SCTS software developed for SMRI.):

$$m\left(C_p\dot{T} + \frac{1}{\rho^2}T\frac{\partial \rho}{\partial T}\bigg|_P \dot{P}\right) = \int_{\partial\Omega} -K\frac{\partial T_{salt}}{\partial n}da + L\dot{C} + <\dot{m}> C_p\left(T_{inj} - T\right) \qquad (1)$$

where $\rho$, $m$, $T$, $P$ are gas density, mass, absolute temperature and pressure, respectively. The gas-state equation can be written $P = \rho rTZ(P,T)$, where $Z = Z(P, T)$ is the gas compressibility factor. Kinetic energy is neglected. The left-hand side reflects the changes in internal energy minus the work of the external forces.

The right-hand side is the sum of the heat flux crossing the cavern walls (discussed later) plus the heat generated during condensation and vaporization of water vapor (This term is neglected in this paper.) plus the enthalpy flux that enters the cavern during gas injection. ($T_{inj}$ is the temperature of the injected gas.) This flux vanishes when no gas is injected in the cavern, $\dot{m} < 0$ (withdrawal) or $\dot{m} = 0$ (standstill).

In the following, simplifying assumptions are made. Only gas withdrawal is considered, $\dot{m} < 0$.

The gas compressibility factor is $Z = 1$ and the gas state equation can be written $P = \rho rT$. Because cavern compressibility is quite small when compared to gas compressibility, the cavern volume is assumed to remain constant. The case of an idealized spherical cavern, radius $a$, is considered.

### 4.3. Gas Withdrawal

Equation (1) then can be re-written

$$m(t)C_v \dot{T}(t) - \dot{m}(t) rT(t) = -4\pi a K_R \int_0^t \dot{T}(\tau) \left[1 + \sqrt{\frac{t_c}{(t-\tau)}}\right] d\tau \qquad T(t=0) = T_0 \quad (2)$$

$t_c = a^2 / \pi k_R$, $k_R$ is the rock thermal diffusivity ($k_R = 3 \times 10^{-6}$ m$^2$/s is typical.), and $r = C_p - C_v$ is the difference between the specific heats of the gas at constant pressure and constant volume, respectively.

### 4.4. Example

When adiabatic evolutions are assumed, the right-hand side of (2) can be neglected, and $T/T_0 = (m/m_0)^{\gamma-1}$, where $\gamma = C_p / C_v$. However, temperature evolutions cannot be considered as adiabatic, as the heat flux crossing the cavern walls - the right-hand side of (2) - cannot be neglected. This is illustrated by Figure 9, in which temperature and pressure evolutions during a cycle are shown (Heath and Benefield, 2010).

The "adiabatic" evolution was added to the initial figure: it is clear that heat transfer from the rock mass cannot be neglected. The significance of the heat transfer depends on withdrawal rate and cavern size. The faster the withdrawal rate, the closer the evolution is to a perfectly adiabatic evolution. For a given relative withdrawal rate (in % per day), the larger the cavern, the closer the evolution is to a perfectly adiabatic evolution. The effect of withdrawal rate, the amount of withdrawn gas and cavern size are discussed in Figures 10 and 11. (Cavern *shape* also can be influential, see Krieter, 2011).

Crossley, 1996, describes a withdrawal test performed in a Melville (Canada) cavern. The measured flow rate, cavern pressure and temperature are drawn on Figure 12 (left). Equation (2) was used to compute temperature evolutions (right). The following values were selected: $\gamma = 1.27$, and $C_{gn}^P = 2347 \ J/kg-K$. The cavern volume is $V = 46\ 153$ m$^3$, but its shape was unknown and the surface/volume area was (artificially) increased by a factor of 2 to reach a good fit between measured and computed values.

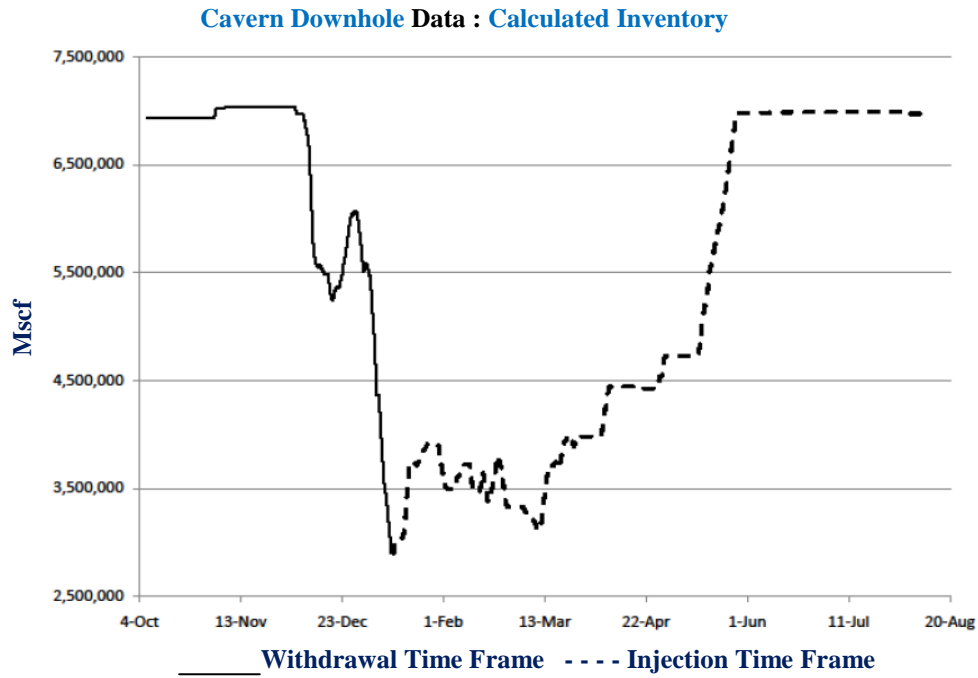

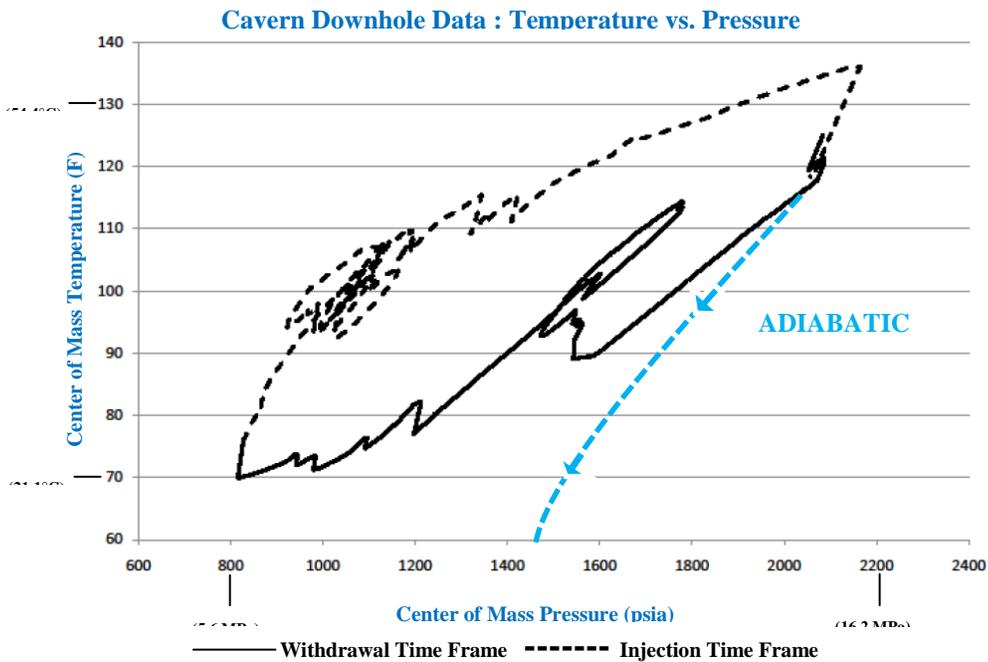

**Figure 9.** Pressure cycle in a Gulf Coast Cavern. Cavern inventory in Mscf (upper picture), gas pressure in psi and temperature in °F (lower picture). Gas withdrawal is represented by the continuous line. The adiabatic curve was added to the original picture. It can be described by the relation $P/P_0 = (T/T_0)^{1-1/\gamma}$. The adiabatic assumption appears irrelevant, because heat flow from the rock mass warms the cavern gas. (After Heath and Benefield, 2010).

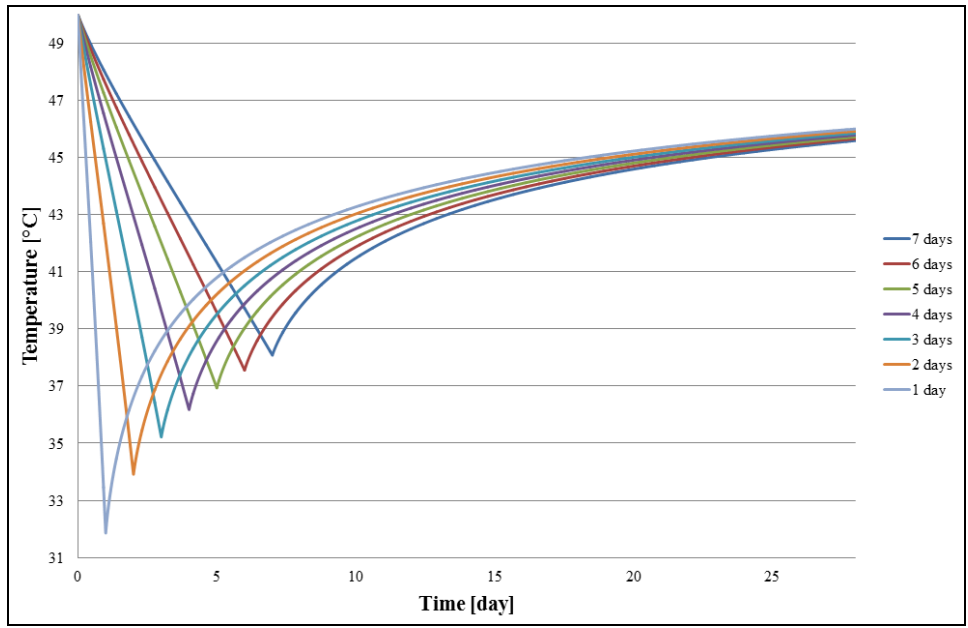

**Figure 10.** Spherical cavern with a volume 216,000 m$^3$: 50% of the initial gas in the cavern is withdrawn over 1, 2, 3, 4, 5, 6 or 7 days. The final adiabatic temperature (when withdrawal is instantaneous) would be –10 °C (14 °F). The actual temperature drop is larger when the withdrawal rate is faster.

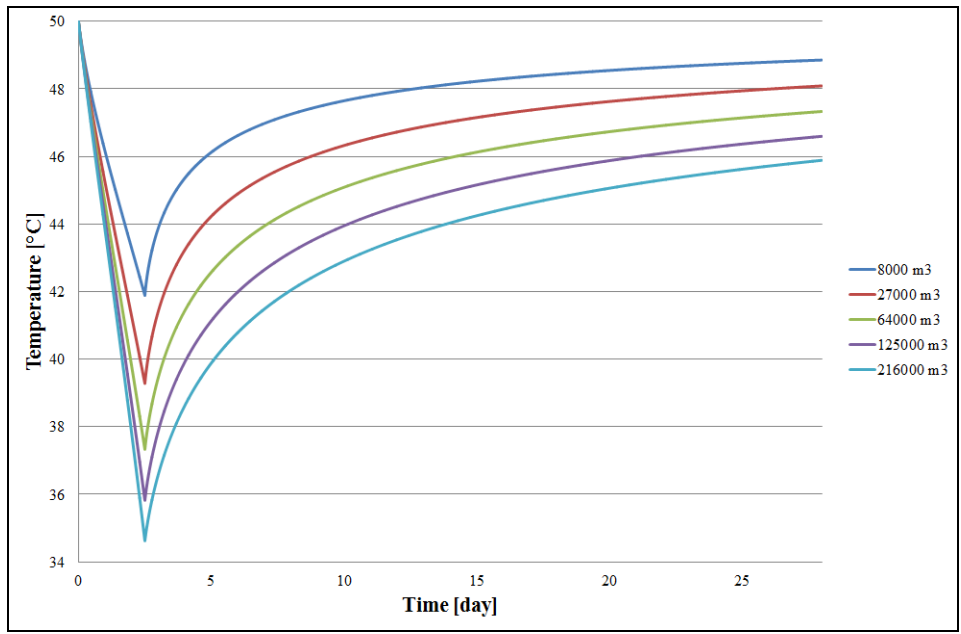

**Figure 11.** Spherical cavern with 25% of the initial gas in the cavern withdrawn over 2 days. The cavern size is 8000 m$^3$, 27,000 m$^3$, 64,000 m$^3$, 125,000 m$^3$ or 216,000 m$^3$ (50,000 bbls, 165,000 bbls, 385,000 bbls, 750,000 bbls or 1,350,000 bbls). The temperature drop is larger when the cavern is larger, as the ratio between heat transfer from the rock mass and the heat capacity of gas is smaller in a larger cavern.

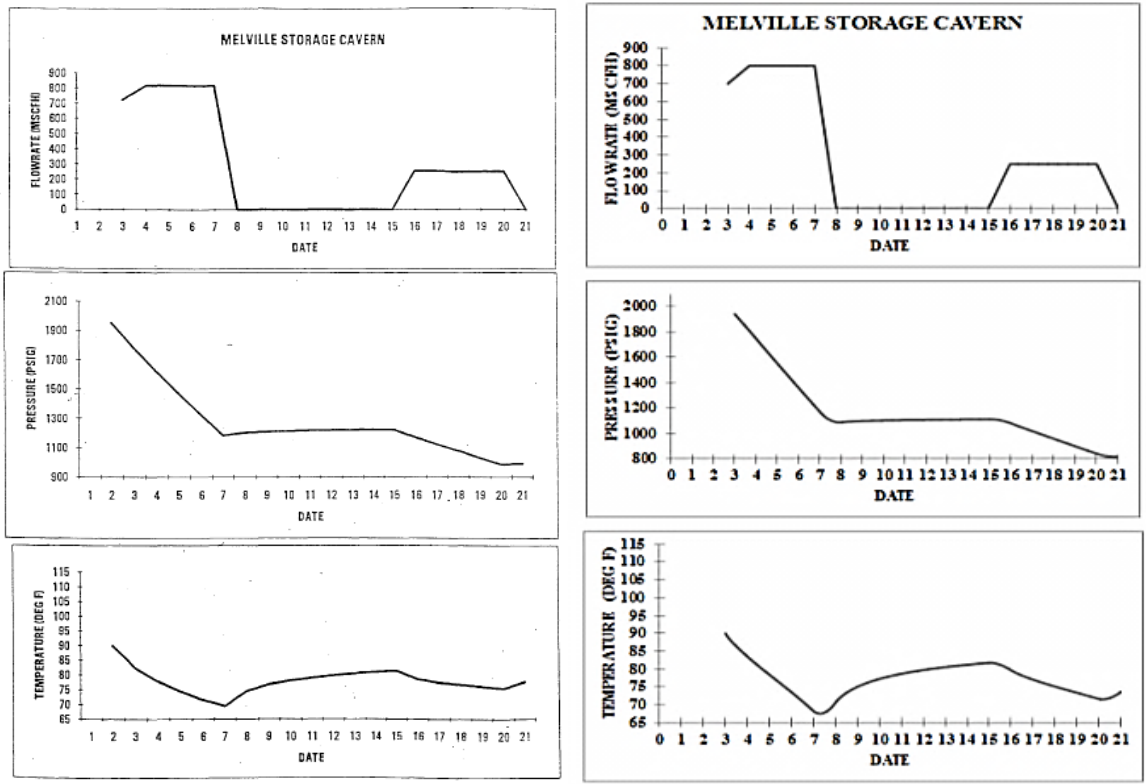

**Figure 12.** Melville Cavern: Withdrawal rate, pressure and temperature evolutions as observed (left) and as computed by LMS (right) (after Crossley, 1996).

## 5. STRESS CHANGES AT CAVERN WALL

### 5.1. Depth of penetration of temperature changes

Solving Equation (2) allows computing the evolution of temperature in the rock mass. Generally speaking, penetration of temperature changes in the rock mass is slow. For instance, when a cold gas temperature was applied at the cavern wall during a $t$-long period of time, the temperature is changed significantly in a domain at a cavern wall with a thickness of $d = (k_R t)^{1/2}$, where $k_R = 3 \times 10^{-6} m^2 / s$ is the thermal diffusivity of salt, or $d \approx 3$ m after $t$ = 1 month. Brouard et al, 2011, considered a spherical cavern and applied a periodic temperature distribution at the cavern wall. The temperature distribution as a function of the distance from the cavern wall when the gas temperature is minimal is shown on Figure 13. Three cycling periods were considered: one day (CAES); one week; and one month (corresponding to an "aggressively" operated HFCGSC). In this last case, temperature fluctuations are divided by a factor of 10 at a 3-m distance from the cavern. (This result does not depend strongly on cavern size.)

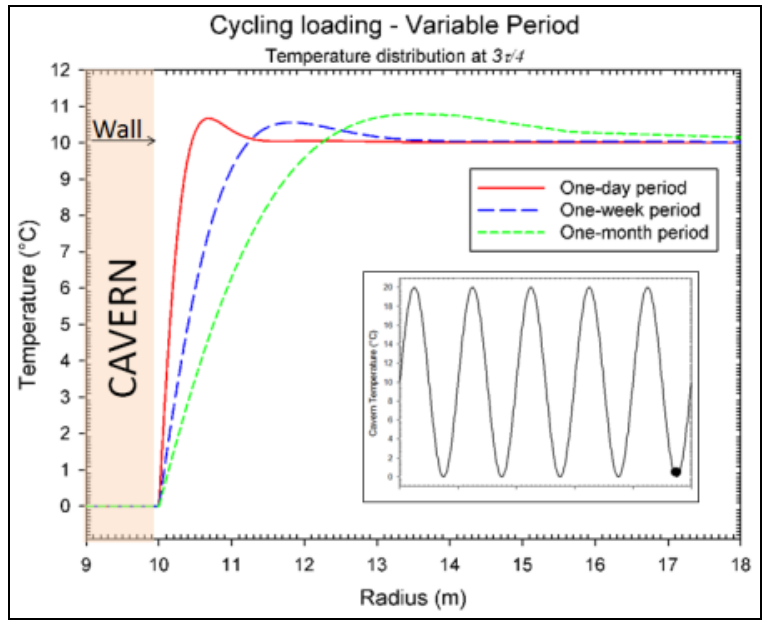

**Figure 13 –** Temperature distribution in the rock mass. Cavern temperature is periodic, varying from 0 °C to 20 °C. Three periods are considered: 1 day, 1 week and 1 month.

### 5.2. "Thermal" stresses

Cooling (or warming) at the cavern wall generates thermal stresses. Orders of magnitudes can be computed easily . Let $\Delta T$ be the difference between the rock temperature at the cavern wall and the virgin temperature (geothermal temperature) at cavern depth. Normal stresses generated by this difference are small (At the cavern wall, the thermal normal stress is zero.); however, tangential thermal stresses are of the order of $\sigma_t = -E\alpha_R \Delta T / (1-\nu)$, $E$ = 18 000 MPa is the elastic modulus, $\nu$ = 0.25 is Poisson's ratio, and $\alpha_R$ = 4 10$^{-5}$/°C is the thermal expansion coefficient. When $\Delta T$ < 0 (gas cooling), the tangential thermal stresses are tensile, and  - $\sigma_t / \Delta T$ = 1 MPa/°C (80 psi/°F). This is a very high figure, as the tensile strength of salt is 1 to 2 MPa. Fracturing is likely. This phenomenon can be explained as follows. At the cavern wall, a relatively thin layer of salt is cooled and should shrink to accommodate temperature drop. In the normal direction, such a contraction is not hindered, and no, or small, thermal stresses appear. This is in sharp contrast with the tangential directions, as contraction of the cold layer at cavern wall is not compatible with the absence of contraction of the deeper layers, which remain warm. For this reason, tangential stresses appear to stretch the superficial layers and balance the temperature-induced contraction.

### 5.3. Fracturing

in many cases, though, tensile stresses are so large that rock tensile strength is exceeded and fractures open. The fractures are perpendicular to the direction of the tensile stresses. Because these stresses are tangential, fractures are perpendicular to the cavern wall. The depth of the fractures must be computed according to the laws of Fracture Mechanics (see Figure 14). However, a simple rule of thumb is that this depth equals the depth of the zone in which stresses (initial stresses + thermal stresses) are tensile.

At the beginning of the process, when the depth of temperature changes is small, the depth of the zone in which stresses are tensile also is small, and a large number of small fractures appear. Salt surface becomes "nacreous and opalizing" (Dreyer, 1977, p .247). When the cold front penetrates deeper into the rock mass, a kind of "fracture selection" takes place. Only a small number of fractures keep on growing, and the distance between two growing fractures becomes larger. At the end of the process in a small cavern, only four fractures remain active (see Figure 2).

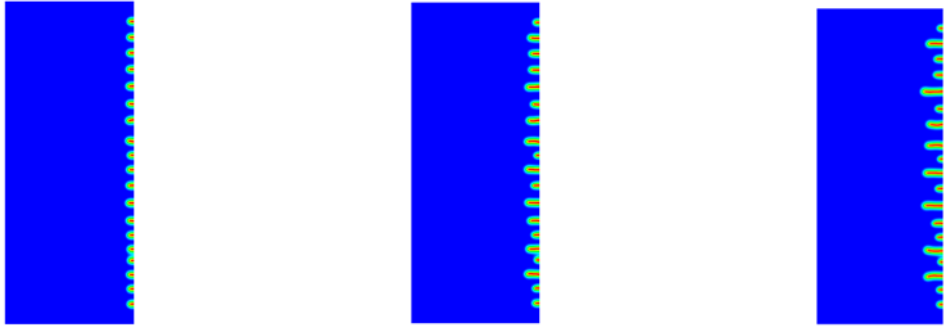

**Figure 14.** Onset of fractures at a cylindrical cavern wall. A temperature drop of *ΔT* = -5 °C (-9 °F) is applied at the cavern wall and kept constant. After some time, fractures appear at the cavern wall. As the cold temperature front penetrates deeper into the rock mass, a smaller number of fractures keep growing. (Computations by P. SicSic and J.J. Marigo, LMS, Ecole Polytechnique).

### 5.4. Fracture aperture

In other words, there is a relation between the distance between two consecutive fractures and the depth of the fractures. The fracture aperture (*e*) also is strongly correlated to the distance between consecutive fractures (*d*) and the average thermal contraction ($\alpha_R \Delta T$):

$$e \approx -d\alpha_R \Delta T$$

For instance, Wallner and Eickemeier, 2001, discussed the onset of fractures in an intake air shaft in a salt formation. "*During the cold season, temperatures in the shaft decreased by 20°C … within a time period of 80 days… horizontal and vertical fractures were detected by routine inspections in the shaft. Theses fractures had an average spacing of about 2.8 m. The fracture aperture amounted up to several mm.*" (p. 365).

This figure is consistent with rough estimations. The thermal expansion coefficient of salt is $\alpha_R = 4 \cdot 10^{-5}$/°C; *ΔT* = - 20°C; and the rock tangential strain generated by rock cooling is $\alpha \Delta T = -0.8 \cdot 10^{-3}$. When fractures are generated every *d* = 2.8 m (9 ft), rock can contract, and the expected fracture aperture is $e \approx d\alpha_R \Delta T = 2$ mm (half an inch). Note that thermal tensile stresses generated at cavern wall can be written $-E\alpha \Delta T /(1-\nu) \approx 15-20$ MPa; they are larger that vertical "initial" stresses at cavern wall (see the Appendix).

### 5.5. Spalling

Conversely, when the rock mass is heated (*ΔT > 0*), instead of cooled, large compressive tangential stresses develop. In general, the normal stress also is compressive, but it is smaller than the two compressive tangential stresses. Such a state of stress is said to be "e*xtensive*". It leads to the formation of multiple flat spalls that fall down to the cavern bottom. When spalls fall, a new cold surface is cleared, and the process goes on. An example of this was given in Figure 1.

Returning to the cooling case, intense tensile stresses generate fractures that open perpendicular to the cavern wall. They delimit plates that remain strongly bonded to the rock mass. It is much less likely that spalls form (than in the heating case), as illustrated by Dreyer's test (Figure 2). Pellizzaro et al., 2011, discussed the stability of such plates and proved that, in most cases, their weight is not large enough to allow plates to break out.

## 5.6. Effective tensile stresses

In Section 5.3 it was assumed that fractures can appear when a tangential tensile stress develops at the cavern wall. (The effect of gas pressure was not taken into account.) However, this criterion may be too optimistic. When performing a hydraulic fracturing test in a borehole, fluid pressure in the borehole is increased to a figure slightly higher than the geostatic pressure to create a fracture. Such tests are performed routinely to assess in-situ stresses. When interpreting such tests, one must take into account the onset of *effective* tensile stresses. Effective stress is the sum of the actual compressive stresses (negative) plus the fluid pressure. When the least-effective stress is tensile (positive), the onset of fracturing is possible. This criterion is much more severe than the no-tensile-stress criterion considered above.

An example of this is provided by Figure 15. Staudtmeister et al, 2010 , computed the three principal stresses (one normal and two tangential) at the wall of a natural gas cavern submitted to pressure cycles. No principal stress is tensile. (The tensile-stress domain is below the red line.) However, it is clear that at days 750, 847 or 887, the cavern pressure is not the least-compressive stress (the blue curve); one of the three *effective* stresses is tensile, and fracturing is possible. Similar results were reached by Dresen and Lux, 2011. Similar computations were performed by Bauer and Sobolik, 2009, Staudtmeister and Zapf, 2010, Karimi-Jafari et al., 2011 and Rokhar et al., 2011. It should be noted that such a phenomenon exists even in a cavern when no thermal effect is considered. Consider the case of a cavern that was kept idle at a low gas pressure over a long period of time (say, several months). It can be proven that during such a period, due to cavern creep closure, deviatoric stress decreases, and tangential stresses progressively become less compressive. When, at the end of such an "idle" period, gas is injected rapidly into the cavern, the gas pressure increases, and *additional* tensile stresses are created. In some cases, effective stresses become tensile, and there is a risk of fracturing (Brouard et al., 2007). However it is believed that this effect might be a "skin" effect (fractures do not penetrate deep into the rock mass) as tensile effective stresses vanish rapidly due to salt creep. Whether such a criterion ("no effective tensile stress") must be taken into account remains unclear.

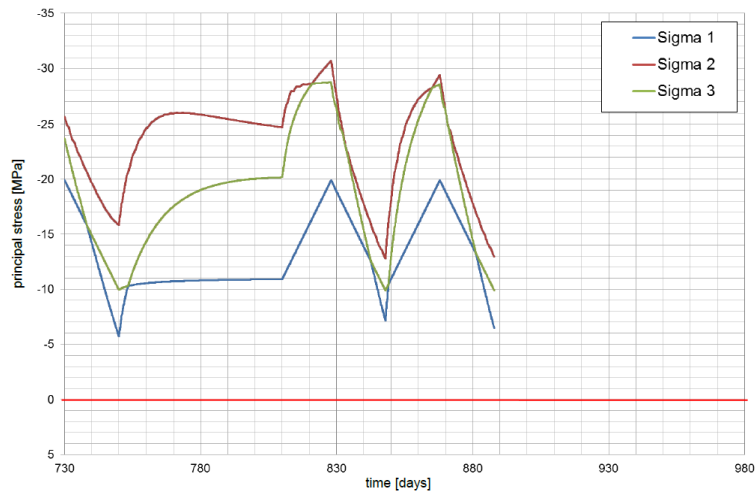

**Figure 15.** Main stresses at the wall of a natural gas cavern submitted to pressure cycles (Staudtmeister et al., 2010).

## 6. CONCLUSION

Rapid gas depressurization leads to gas cooling that is followed by slow gas warming when the cavern is kept idle. The decrease in the temperature of gas depends upon the relative withdrawal rate (in %/day), and cavern size and shape. Gas cooling may result in the onset of tensile stresses at cavern walls and roofs that may generate fractures or cracks. However, in most cases, the depth of penetration of these fractures is small, and they are perpendicular to the cavern wall. The distance between two parallel fractures becomes larger when fractures penetrate deeper into the rock mass, as some fractures stop growing. Fractures form a polygonal pattern. Salt slabs are created, with boundaries formed by the opened fractures. As long as the depth of penetration of the fracture remains small, these slabs remain strongly bonded to the rock mass, and it is believed that, in many cases, their weights are not large enough to allow them to break off the cavern wall.

**Acknowledgements**

This study was funded partially by the French *Agence Nationale de la Recherche* in the framework of the SACRE Project, which includes researchers from EDF, GEOSTOCK, PROMES (Perpignan), HEI (Lille) and Ecole Polytechnique (Palaiseau). Special Thanks to K.Sikora.

**APPENDIX**

In this Appendix, we discuss the stress distribution at the wall of an idealized cylindrical cavern (or shaft) submitted to temperature changes. A simplified solution is proposed.

We assume, first, that before the temperature drop, the cavern has been left idle for a long period of time. Steady-state stress distribution was achieved:

$$\begin{cases} \sigma_{rr}(t=\infty) = -P_\infty + (P_\infty - P_0)(a/r)^{2/n} \\ \sigma_{\theta\theta}(t=\infty) = -P_\infty + \left(1 - \frac{2}{n}\right)(P_\infty - P_0)(a/r)^{2/n} \\ \sigma_{zz}(t=\infty) = -P_\infty + \left(1 - \frac{1}{n}\right)(P_\infty - P_0)(a/r)^{2/n} \end{cases} \quad (3)$$

$P_\infty$ is the geostatic pressure at cavern depth; $P_\infty = \gamma_R z$, $\gamma_R = 0.95$ psi/ft or 0.022 MPa/m is typical.

$P_0$ is the gas pressure, $P_0 = 0$ in a mine shaft.

*a* is the cavern radius.

*n* is the exponent of the Norton-Hoff power law ($\dot{\varepsilon} = A\sigma^n$, *n* = 3 to 6).

Now, it is assumed that the gas in the cavern is cooled by $\theta_a = \theta_a(t) < 0$, resulting in a change in rock temperature by $\theta = \theta(r,t) < 0$. Additional temperature-induced stresses can be written:

$$\begin{cases} \Delta'\sigma_{rr}(r,t) = -(a/r)^2 \dfrac{E\alpha}{1-v} \int_a^r u\theta(u,t)du \\ \Delta'\sigma_{\theta\theta}(r,t) = +(a/r)^2 \dfrac{E\alpha}{1-v} \int_a^r u\theta(u,t)du - \dfrac{E\alpha\theta(r,t)}{1-v} \\ \Delta'\sigma_{zz}(r,t) = -\dfrac{E\alpha\theta(r,t)}{1-v} \end{cases} \quad (4)$$

where $E$ and $v$ are the salt elastic modulus and Poisson's ratio, respectively, and $\alpha$ is the thermal expansion coefficient of salt. Typical values are $E$ = 18 000 MPa, $v$ = 0.25, and $\alpha$ = 4 $10^{-5}$/°C. The order of magnitude of the thermal stresses at the cavern wall is $E\,\alpha/(1-v)$ = 1 MPa/°C, or 80 psi/°F. As rock is cooled ($\theta_a = \theta_a(t) < 0$), these stresses are tensile stresses. (They can open fractures.) 1 MPa/°C is a large figure, as the tensile strength of salt is 1 to 2 MPa.

In many cases, gas cooling results from gas depressurization, and pressure change must be taken into account because it generates additional stresses, which, to some extent, mitigate the mechanical effects of gas cooling. When gas pressure drops by $\Delta P > 0$, the cavern pressure is $P_0 - \Delta P$, and additional stresses are generated:

$$\begin{cases} \Delta\sigma_{rr}(r) = +\Delta P\,(a/r)^2 \\ \Delta\sigma_{\theta\theta}(r) = -\Delta P\,(a/r)^2 \\ \Delta\sigma_{zz}(r) = 0 \end{cases} \quad (5)$$

When gas cooling (or gas depressurization) is fast enough, the initial visco-plastic steady-state stresses are not left enough time to redistribute. Stresses at cavern wall ($r = a$) can be written

$$\begin{cases} \sigma_{rr}(t=\infty) = -P_0 + \Delta P \\ \sigma_{\theta\theta}(t=\infty) = -P_\infty + \left(1 - \dfrac{2}{n}\right)(P_\infty - P_0) - \dfrac{E\alpha\theta_a}{1-v} - \Delta P \\ \sigma_{zz}(t=\infty) = -P_\infty + \left(1 - \dfrac{1}{n}\right)(P_\infty - P_0) - \dfrac{E\alpha\theta_a}{1-v} \end{cases} \quad (6)$$

In the case of a mine shaft, P0 = ΔP = 0, and

$$\begin{cases} \sigma_{rr}(t=\infty) = 0 \\ \sigma_{\theta\theta}(t=\infty) = -\left(\dfrac{2}{n}\right)P_\infty - \dfrac{E\alpha\theta_a}{1-v} \\ \sigma_{zz}(t=\infty) = -\left(\dfrac{1}{n}\right)P_\infty - \dfrac{E\alpha\theta_a}{1-v} \end{cases} \quad (7)$$

This very simple model predicts that horizontal fractures can appear at a shaft wall when $\theta_a < -\dfrac{(1-v)P_\infty}{E\alpha n}$.